\documentclass{iaus}
\usepackage{graphicx}
\usepackage{natbib}

\title[OB runaways from the stellar disk(s) in the Galactic center] 
{Getting a kick out of the stellar disk(s) in the galactic center}

\author[Perets, Kupi \& Alexander]   
{Hagai B. Perets, G\'abor Kupi \& Tal Alexander}

\affiliation{Weizmann Institute of Science, POB 26, Rehovot 76100, Israel.}

\pubyear{2007}
\volume{246}  
\pagerange{##--##}
\date{?? and in revised form ??}
\jname{IAU Symp. 246 - Evolution of dense stellar systems}
\editors{E. Vesperini (Chief Editor), M. Giersz, A. Sills}
\begin{document}

\maketitle

\begin{abstract}
Recent observations of the Galactic center revealed a nuclear disk of young OB stars, in addition to many similar outlying stars with higher eccentricities and/or high inclinations relative to the disk (some of them possibly belonging to a second disk). Binaries in such nuclear disks, if they exist in non-negligible fractions, could have a major role in the evolution of the disks through binary heating of this stellar system. We suggest that interactions with/in binaries may explain some (or all) of the observed outlying young stars in the Galactic center. Such stars could have been formed in a disk, and later on kicked out from it through binary related interactions, similar to ejection of high velocity runaway OB stars in young clusters throughout the galaxy.
\end{abstract}

Recent observations have revealed the existence
of many young OB stars in the galactic center (GC). Accurate measurements
of the orbital paramters of these stars give strong evidence for the
existence of a massive black hole (MBH) which govern the dynamics
in the GC \citep{eis+05}. Most of the young stars are observed to be OB stars in the central 0.5 pc around the MBH. Many of them are observed in a coherent stellar disk or two perpendicular disks configurations \citep{lu+06,pau+06}. Others are observed to be have inclined and/or eccentric ($>0.5$) orbits relative to the stellar disks (hereafter outliers). The inner 0.04 pc near the MBH contain only young B-stars, that possibly have a different origin (e.g. \citealt{lev07,per+07}). 

It was suggested that the disk stars have been formed
 a few Myrs years ago in a fragmenting gaseous disk \citep{nay+05b,lev07}.
However, the origin of outliers from the disk is difficult to explain in this way. These stars are observed to have very similar stellar properties
to the young disk stars (types, lifetimes), but have more inclined and/or eccentric orbits. Many suggestions have been made for the origin of these stars
 \citep[and references therein]{mil+04,pau+06,ale+07,yu+07}.
Here we suggest a different process in which young stars in the GC stellar
disks were kicked into high inclinations and/or eccentricities, in a similar way to OB runaway stars ejected from open clusters. Such a 
scenario could explain some of the puzzling orbital properties
of the young stars in the GC. 

A considerable fraction of the early OB stars in the solar neighborhood 
have large peculiar velocities ($40\le v_{pec}\le200$ km s$^{-1}$; e.g. \citealt{hoo+01}) and are observed in isolated locations; these are the so-called runaway stars \citep{bla61}.
\begin{figure}
\includegraphics[clip,scale=0.3]{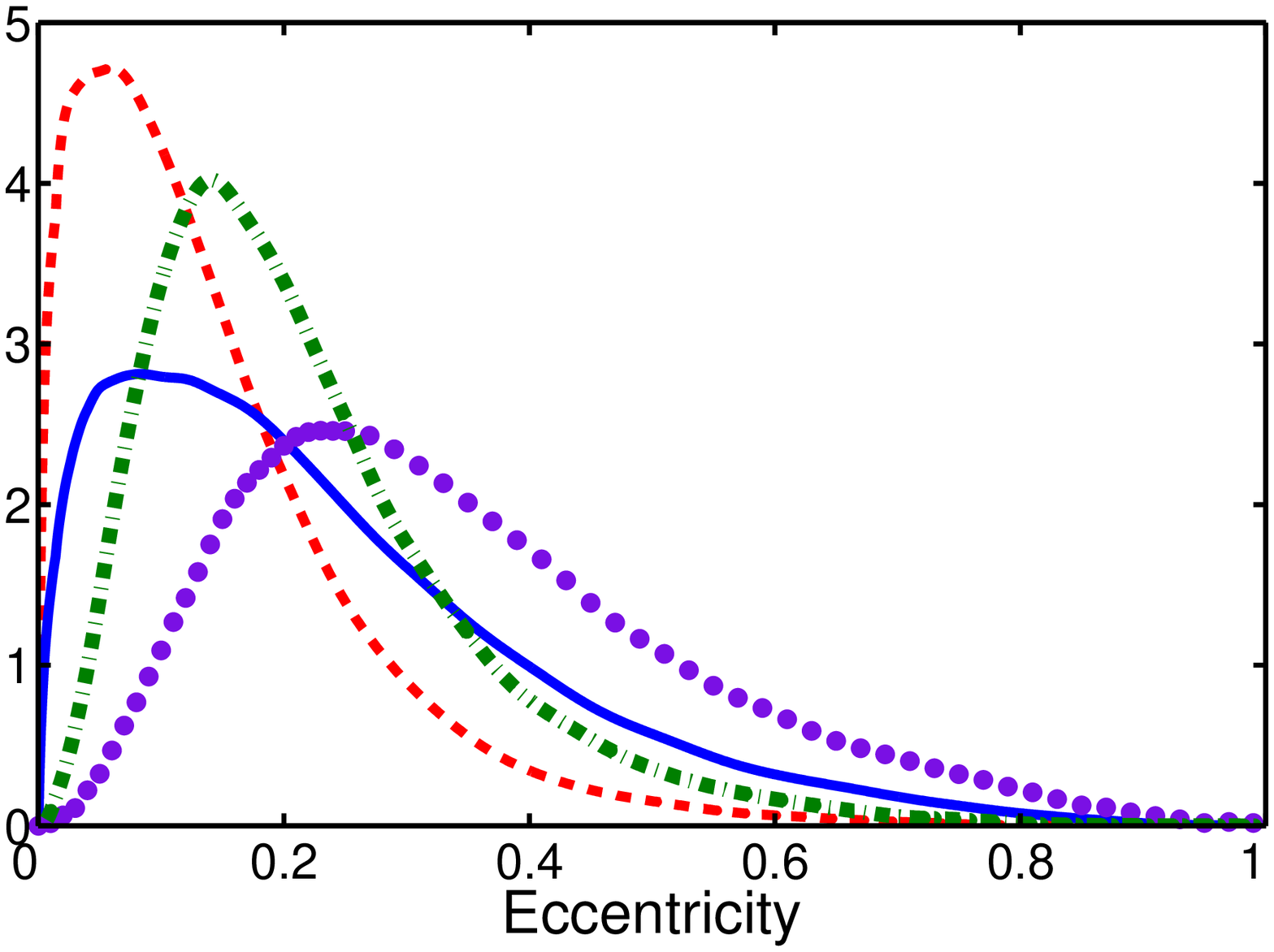}
\includegraphics[clip,scale=0.3]{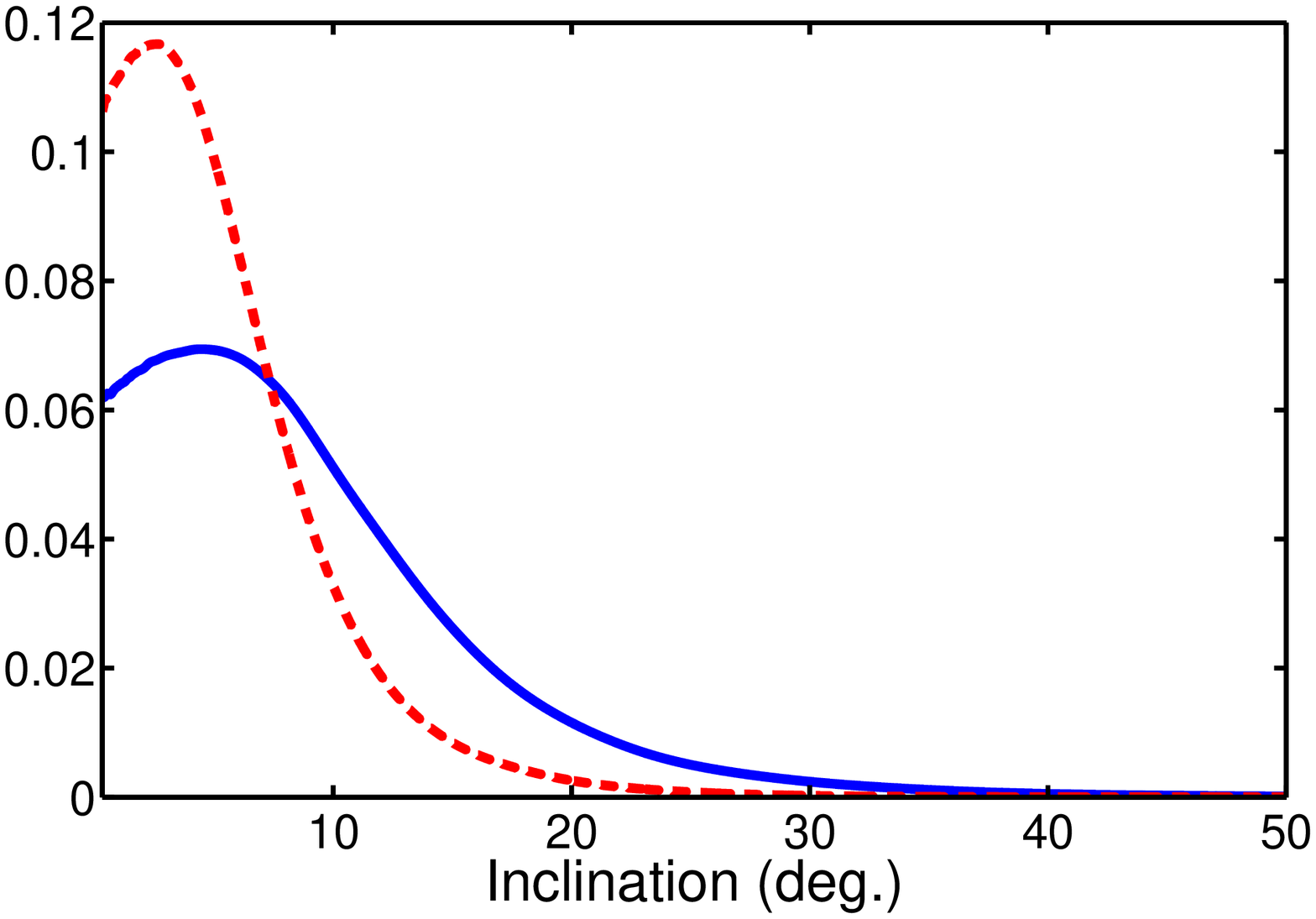}
\caption{\label{f:eccs} Left: Eccentricity distribution of outliers in the Galactic center.
Monte-Carlo results assuming different models for the Maxwellian
 kick velocity distributions; isotropic, $v\sim 100$ km s$^{-1}$ (solid)
, isotropic, $v\sim 60$ km s$^{-1}$ (dashed), planar (co-planar with the disk), $v\sim 100$ km s$^{-1}$ (dotted)  and planar,  $v\sim 60$  km s$^{-1}$ (dash-dotted).
Right: Inclination distribution of outliers in the Galactic center. Monte-Carlo results assuming different models for the Maxwellian
 kick velocity distributions; isotropic, $v\sim 100$ km s$^{-1}$ (solid)
, isotropic, $v\sim 60$ km s$^{-1}$ (dashed).}
\end{figure}
Two mechanisms are thought to eject OB runaway stars, both involve binarity (or higher multiplicity). In the binary
supernova scenario (BSS; \citealt{bla61}) a star is kicked at high velocity 
following a supernova explosion of its companion. 
In the dynamical ejection scenario (DES; \citealt{pov+67})
runaway stars are ejected through gravitational interactions between
stars in dense, compact clusters. The DES is more likely for the disk stars, given the short lifetime of the disk.

The binary properties of stars in the GC are unknown, but observations 
of eclipsing binaries in the GC suggests they are not fundamentally different from that observed in the solar neighborhood \citep{dep+04,mar+06,raf+07}. 
The conditions in the stellar disks in the GC are highly favorable for the DES given the high stellar densities ($>10^6$ pc$^{-3}$), the low velocity 
dispersion and the masses of the stars in
the stellar disk.
It is thus quite likely that dynamical ejection in the disk is more frequent 
and efficient than in normal Galactic star forming regions. In order to escape the GC cusp, a disk star initially bounded to the MBH
needs a kick velocity of a few$\times10^{2}$ km s$^{-1}$.
Most of the observed OB runaways in the galaxy do not reach such
high velocities. Therefore runaways from the stellar disk would not escape, and would remain in the cusp. However, such kicks could considerably change the orbits of these stars. In fig. 1 we show the inclination and eccentricity distribution of OB stars that were formed in a disk on circular orbits and were kicked out of the disk with velocities typical of OB runaways ($40-200$ km s$^{-1}$).
Stars in the disk with near circular Keplerian orbits, would
be kicked into more eccentric and inclined orbits (on average). Their orbits 
should also display some correlations between inclination, eccentricity and mass. Such ``failed'' runaways could possibly explain the outliers from and inside the disks in the GC, which are difficult to explain otherwise.

\end{document}